\documentclass[lettersize,journal]{IEEEtran}
    \usepackage{booktabs}
    \usepackage{diagbox}
    \usepackage{makecell}
    \usepackage{array}
    \usepackage{graphicx,amssymb,amsmath}
    \usepackage{multicol}
    \usepackage[noadjust]{cite} 
    \usepackage{setspace}
    \usepackage{subfigure}
    \usepackage{tabularx}
    \usepackage{graphicx}
    \usepackage{float}
    \usepackage {url}
    \usepackage{stfloats}
    \usepackage{amsthm,pifont}
    \usepackage{flushend}
    \usepackage{cases,subeqnarray}
    \usepackage{bm,multirow,bigstrut}
    \usepackage{amsmath, amsthm, amssymb}
    \usepackage{textcomp}
    \usepackage{latexsym,bm}
    \usepackage{booktabs}
    \usepackage{mathtools}
    \usepackage{dsfont}
    \usepackage{extarrows}
    \usepackage{epsfig}
    \usepackage{epstopdf}
    \usepackage[table]{xcolor}
    \usepackage{colortbl} 
    \usepackage{multirow, hhline}
    \usepackage{xcolor}
    \usepackage{array} 
    \usepackage{algorithmicx,algorithm}
    \usepackage{hyperref}

    \theoremstyle{plain}

    \theoremstyle{plain}

    \usepackage{amsmath}

    \IEEEoverridecommandlockouts
    \begin{document}

\title{Knowledge-driven Reasoning for Mobile Agentic AI: Concepts, Approaches, and Directions}
\author{Guangyuan Liu, Changyuan Zhao, Yinqiu Liu, Dusit Niyato,~\IEEEmembership{Fellow,~IEEE}, and Biplab Sikdar,~\IEEEmembership{Fellow,~IEEE} 

\thanks{G.~Liu is with the College of Computing and Data Science, the Energy Research Institute @ NTU, Interdisciplinary Graduate Program, Nanyang Technological University, Singapore (e-mail: liug0022@e.ntu.edu.sg).}
\thanks{Y.~Liu, C.~Zhao, D. Niyato are with the College of Computing and Data Science, Nanyang Technological University, Singapore (e-mail: yinqiu001@e.ntu.edu.sg, zhao0441@e.ntu.edu.sg, dniyato@ntu.edu.sg).}
\thanks{B.~Sikdar is with the Department of Electrical and Computer Engineering, National University of Singapore, Singapore(e-mail: bsikdar@nus.edu.sg).}
}
\maketitle

\begin{abstract}
Mobile agentic AI is extending autonomous capabilities to resource-constrained platforms such as edge robots and unmanned aerial vehicles (UAVs), where strict size, weight, power, and cost (SWAP-C) constraints and intermittent wireless connectivity limit both on-device computation and cloud access. Existing approaches mostly optimize per-round communication efficiency, yet mobile agents must sustain competence across a stream of tasks. We propose a knowledge-driven reasoning framework that extracts reusable decision structures from past execution, synchronizes them over bandwidth-limited links, and injects them into on-device reasoning to reduce latency, energy, and error accumulation. A DIKW-inspired taxonomy distinguishes raw observations, episode-scoped traces, and persistent cross-task knowledge, and categorizes knowledge into retrieval, structured, procedural, and parametric representations, each with a distinct tradeoff between reasoning speedup and failure risk. A key finding is that knowledge exposure is non-monotonic: too little forces costly trial-and-error replanning, while too much introduces conflicting cues and errors. A UAV case study validates the framework, where a compact knowledge pack synchronized over intermittent backhaul enables a 3B-parameter onboard model to achieve perfect mission reliability with lower reasoning cost than both knowledge-free on-device reasoning and cloud-centric replanning.
\end{abstract}
\section{Introduction}
\label{sec:introduction}

Agentic AI is transitioning from cloud-centric deployment to mobile platforms, such as smartphones, edge robots, and unmanned aerial vehicles~\cite{memory_survey_agents_2025}. Different from cloud agents that can rely on abundant compute and stable access to large external memories, mobile agents must complete long-horizon perceive--reason--act loops under strict size, weight, power, and cost (SWAP-C) constraints and intermittent wireless connectivity~\cite{Survey_2512}. Consequently, a mobile agent must decide not only what to infer, but also what to store locally, what to retrieve opportunistically, and what to transmit under bandwidth and energy limits.

Recent progress has improved communication efficiency for mobile intelligence, especially through semantic communication, where each interaction transmits a compressed representation of current observations (such as tokens, latents, or patches) to reduce per-round bandwidth while preserving task-relevant content. However, per-round compression primarily optimizes a single exchange, whereas mobile autonomy is inherently sequential: the agent executes a stream of tasks over time and must remain competent as goals, environments, and user preferences evolve. Consequently, the system bottleneck shifts from one-shot transmission to sustained competence under cumulative resource constraints. Rather than repeatedly spending bandwidth and energy to reconstruct task-local context for each new task, a mobile agent should identify reusable structure that benefits many future tasks and communicate it in a way that amortizes cost over time.

We refer to this paradigm as knowledge communication, which supports a transition from data-driven signal exchange to knowledge-driven capability evolution. In this context, knowledge broadly encompasses transferable structures, such as mappings, priors, templates, or policies distilled from accumulated task execution, and it can exist in both implicit (parametric) and explicit (externalizable) forms~\cite{farahani_interplay_2024}.

To operationalize this perspective, we adopt a Data-Information-Knowledge-Wisdom (DIKW)-inspired taxonomy and reinterpret it for agentic systems~\cite{rowley2007wisdom}. Data corresponds to raw observations. Information corresponds to task-grounded, episode-scoped traces and outcomes produced during execution, such as interaction trajectories, intermediate artifacts, tool traces, and task outcomes. Context is a minimal working pack constructed from information to condition inference within a single task. In contrast, knowledge is a reusable and persistent asset distilled from accumulated execution information, designed to support cross-task transfer so that its acquisition and communication cost can be amortized over time. This time-scale separation further implies a separable knowledge lifecycle, including knowledge extraction (or promotion), knowledge communication and synchronization, and knowledge utilization through activation during reasoning.

A central challenge is that knowledge influences not only accuracy, but also the trajectory-level cost of reasoning. For instance, when a UAV aerial base station must replan its serving order after a runtime disruption, each additional reasoning step consumes tokens, on-device memory, and potentially triggers signaling over a constrained backhaul link. More generally, in mobile agentic pipelines, reasoning unfolds as a multi-step trajectory whose cost accumulates through token generation and compute, memory and history retention, as well as branching and backtracking under uncertainty. Knowledge can reshape this trajectory through multiple representations, including retrieval knowledge for instance reuse, structured knowledge for constraint-aware composition, procedural knowledge for deterministic workflow execution, and parametric knowledge for low-latency pattern completion. Importantly, knowledge exposure is non-monotonic: insufficient effective knowledge leads to search-heavy reasoning, whereas excessive or weakly aligned knowledge can introduce distraction, branching, and hallucination risk. Therefore, the key system problem is not maximizing stored knowledge volume, but controlling relevance-aware activation under tight communication and energy budgets. In wireless-native deployments, this activation control is further coupled with link availability and channel quality, since knowledge must be synchronized over intermittent and capacity-limited backhaul before it can be utilized onboard.

The contributions of this paper are summarized as follows:
\begin{itemize}
    \item We introduce a DIKW-inspired operational taxonomy for mobile agentic systems and clarify the distinction among data, information, context, and knowledge by their roles and lifetimes under SWAP-C and wireless constraints.
    \item We formalize knowledge-driven reasoning mechanics by modeling reasoning as a cost-bearing trajectory and by categorizing knowledge into retrieval, structured, procedural, and parametric representations, highlighting their distinct effects on trajectory length, branching, and failure modes.
    \item We reveal a non-monotonic tradeoff in knowledge activation and validate it through a UAV aerial base station case study under intermittent backhaul and runtime disruptions, where a compact, cacheable knowledge pack enables an onboard model to achieve perfect mission reliability with lower reasoning cost.
\end{itemize}

\section{Foundations of Knowledge-Driven Reasoning: Concepts and Representations}
\label{sec:concepts_representations}

\begin{figure*}[t]
    \centering
    \includegraphics[width=0.98\textwidth]{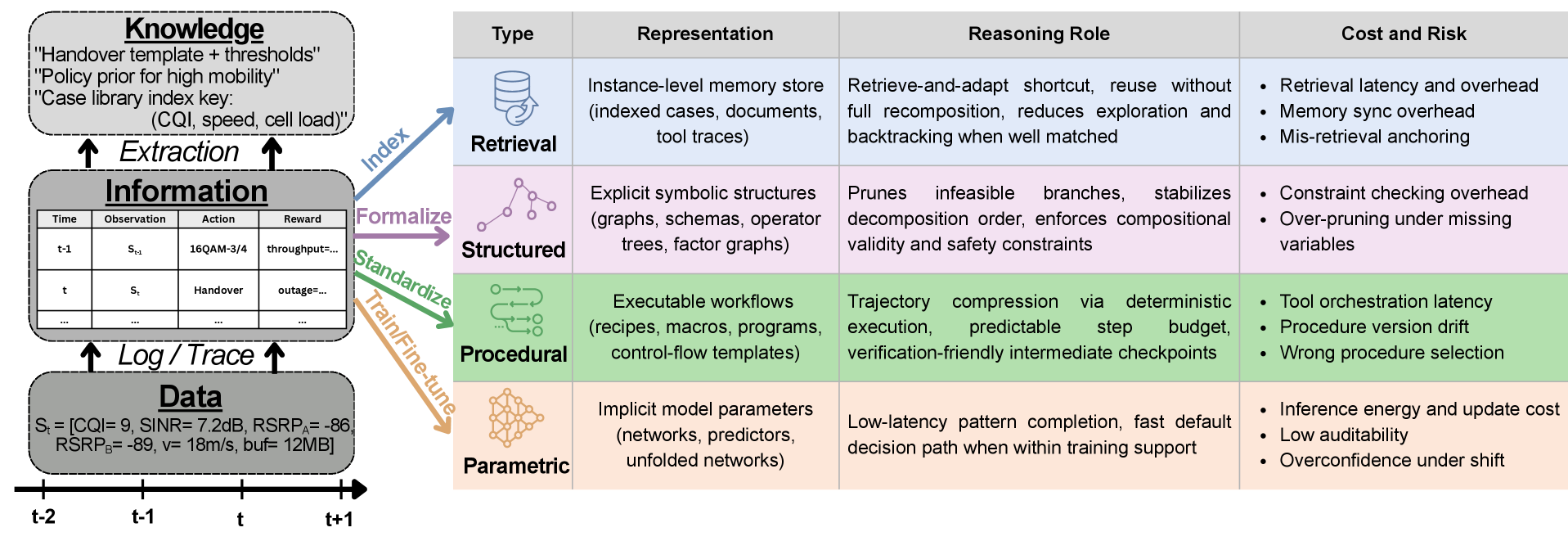}
    \caption{Operational foundations of knowledge-driven reasoning for mobile agents. A DIKW-inspired time-scale hierarchy is grounded in a representative wireless mobility episode, where per-round observations constitute data, episode-scoped execution traces constitute information, and reusable structures promoted from information constitute knowledge. Four knowledge representations are shown with their operational roles, system costs/risks, and extraction routes: indexing (retrieval), formalization (structured), standardization (procedural), and parameter update (parametric).}

    \label{fig:knowledge_reasoning}
\end{figure*}

In this section, we present operational foundations for knowledge-driven reasoning in mobile agentic AI under SWAP-C and wireless constraints. We first define reasoning as a multi-step trajectory in agentic pipelines, clarifying what constitutes intermediate states and why system cost accumulates along multi-step inference and tool use. We then distinguish data, information, and knowledge through a DIKW-inspired time-scale separation, and categorize knowledge into four representations (retrieval, structured, procedural, and parametric) together with their major system costs and risks. Fig.~\ref{fig:knowledge_reasoning} summarizes this foundation using a representative wireless mobility episode (e.g., handover).

\subsection{Basic Concepts: Reasoning and Knowledge in Agentic Pipelines}
\label{subsec:basic_concepts}

In agentic AI, reasoning refers to the process that transforms task inputs into decisions or outputs through a sequence of intermediate states. Concretely, given a task specification (e.g., a query, a goal, and optional constraints) and available resources (e.g., tools and memories), an agent produces an output by iteratively forming hypotheses, selecting operations, and updating its internal state. Depending on the agent design, these intermediate states may be realized as natural-language deliberations, structured plans, executable programs, tool call arguments, retrieved evidence, or verification signals~\cite{reasoningforwireless,gao2025efficient}.

Current agentic AI systems implement this process through several recurring paradigms, such as chain-style deliberation that generates explicit intermediate steps~\cite{reasoningforwireless}, tool-augmented reasoning that alternates between internal reasoning and external actions, search-oriented reasoning that expands multiple candidate paths, and program-aided reasoning that converts deliberation into executable code~\cite{gao2025efficient}. Although these paradigms differ in form, they share a common structure, i.e., reasoning unfolds as a multi-step trajectory, and system cost accumulates along that trajectory.

To define knowledge in a way that is compatible with the trajectory view, we adopt a DIKW-inspired hierarchy and reinterpret it for agentic reasoning systems~\cite{rowley2007wisdom}. In repeated wireless communication tasks (such as link adaptation and handover), data are raw observations, information consists of episode-scoped traces and outcomes produced during execution, and knowledge is a reusable and persistent asset distilled from accumulated execution information to support cross-task transfer, so that its acquisition and communication cost can be amortized over time. This hierarchy further suggests a separable knowledge lifecycle in knowledge-driven reasoning, including knowledge extraction (or promotion), knowledge communication and synchronization, and knowledge utilization through relevance-aware activation that injects a budgeted subset of knowledge into the ongoing reasoning trajectory.


\subsection{Knowledge Representations and Operational Capabilities}
\label{subsec:knowledge_representations}


With these definitions in place, we next categorize knowledge by representation. Knowledge is often discussed as if it were a single quantity that can be increased or decreased, yet this abstraction is insufficient for agentic reasoning because knowledge differs not only in content but also in form. Representation determines how knowledge is accessed, how it interacts with reasoning, and what system costs it introduces. Therefore, we organize the following discussion around four representations, each emphasizing a distinct operational capability.

\subsubsection{Retrieval Knowledge}
\label{subsubsec:retrieval_knowledge}
Retrieval knowledge is stored as instance-level records, such as text chunks, indexed documents, tool traces, or solved examples, and is accessed through semantic or vector retrieval mechanisms~\cite{asai2024self}. Its core reasoning role is direct reuse without task recomposition. In wireless communication, this corresponds to retrieving past episodes with similar radio and mobility context, such as comparable channel quality indicator (CQI), signal-to-interference-plus-noise ratio (SINR), reference signal received power (RSRP), and reference signal received quality (RSRQ), together with the associated actions (such as modulation and coding scheme (MCS) selection, beam switching, and handover triggers) and their observed outcomes. When a task closely matches a stored instance, the agent can retrieve the record and reuse it as a reference or even as a near-complete solution, thereby bypassing much of the intermediate reasoning trajectory.

This representation is particularly attractive in mobile settings because it can reduce computation and latency through near-direct reuse. However, this efficiency is accompanied by two inherent weaknesses that are critical from a system perspective. First, retrieval knowledge is coverage-limited, since its effectiveness depends on the existence of sufficiently similar instances in the memory or database. Second, it is precision-sensitive, because misaligned retrievals can anchor the agent to inappropriate solutions (such as an overly aggressive MCS under shifted interference) and produce high-confidence but incorrect outputs.

\subsubsection{Structured Knowledge}
\label{subsubsec:structured_knowledge}
Structured knowledge is represented as explicit symbolic forms, such as operator trees, equations, graphs, or constraint networks~\cite{han2025reasoning}. Rather than encoding a ready-to-execute solution for a specific task instance, it encodes the structural form of a task class, namely how components are composed and which relations must hold among them. In wireless communication, this structure can be instantiated as explicit constraints and relational models, such as feasibility rules for handover triggering, reliability constraints linking link quality to admissible rate choices, or interference coupling graphs among links. Typical formats include parse trees for equations, abstract syntax trees for programs, and factor graphs for constrained inference.

This representation is particularly important when tasks exhibit strong compositional structure, such as multi-step tool programs, symbolic transformations, or planning under constraints. From a mobile systems perspective, structured knowledge is appealing because it is compact, interpretable, and naturally reusable across tasks. However, it typically does not specify a complete execution path. Instead, it acts as a scaffold that stabilizes exploration by pruning invalid options early and preparing the agent for subsequent trajectory compression through procedural activation or controlled search.

\subsubsection{Procedural Knowledge}
\label{subsubsec:procedural_knowledge}
Procedural knowledge is represented as an executable procedure, such as a stepwise recipe, a control-flow template, or a reusable tool macro. Rather than specifying what is valid, it specifies what to do next. Accordingly, its core reasoning role is trajectory compression through deterministic execution. In wireless communication, procedural knowledge corresponds to standardized decision workflows, such as a measurement aggregation step followed by candidate evaluation, decision, and verification for beam switching or handover triggering. Once an appropriate procedure is activated, the agent can replace open-ended exploration with a short sequence of prescribed operations, thereby reducing branching, backtracking, and intermediate-state accumulation.

This representation is particularly effective for mobile agentic AI because it yields predictable and budget-friendly reasoning. Since the execution path is short and stable, the system can better budget token generation, memory access, and tool invocation under SWAP-C constraints, while also simplifying verification because intermediate stages are well defined. Moreover, procedural assets are naturally reusable across tasks, especially when tasks share common subroutines, such as perception-to-action tool chains, planning primitives, or standardized transformation patterns.

However, procedural knowledge shifts the bottleneck from derivation to selection. If the agent activates an incorrect procedure, or activates a correct procedure under incompatible assumptions (such as applying a low-mobility handover procedure in a high-mobility regime), the resulting execution may fail quickly or, more critically, produce plausible but incorrect outputs. Therefore, effective use of procedural knowledge requires relevance-aware activation and lightweight validity checks that confirm applicability before committing to deterministic execution~\cite{nagy2026chopchop}.

\subsubsection{Parametric Knowledge}
\label{subsubsec:parametric_knowledge}
Parametric knowledge is encoded implicitly in the model parameters as distributed representations acquired through pretraining and fine-tuning. Rather than exposing explicit intermediate structure, the model internalizes statistical regularities that enable direct mapping from inputs to outputs. In wireless communication, this often takes the form of a learned policy or predictor that maps radio and mobility observations to actions or performance estimates, enabling fast decision making without explicit retrieval. Concrete examples include trained channel prediction networks that forecast future channel states from historical measurements, and mobility-aware scheduling policies that learn to allocate resources based on user trajectory patterns~\cite{reasoningforwireless}. 
Accordingly, its core reasoning role is low-latency inference with limited transparency. When the input falls within the support of learned patterns, parametric knowledge can collapse a potentially long reasoning trajectory into a short inference process, thereby achieving minimal delay.

From a mobile systems perspective, this representation is attractive because it operates without external retrieval or explicit structured interfaces at inference time, and thus avoids additional communication and memory access overhead. However, its effectiveness degrades under distribution shift, ambiguous task formulations, or safety-critical conditions. In wireless systems, such shift can arise from time-varying interference, mobility, topology changes, and evolving traffic loads, which makes purely parametric shortcuts fragile without external grounding. Since intermediate reasoning states are not necessarily explicit, parametric reasoning offers limited support for verification, auditing, and rollback, which complicates reliability management under tight resource budgets. Consequently, while parametric knowledge provides an efficient baseline in stable regimes, it is typically complemented by retrieval mechanisms, structured constraints, or procedural checks when robustness and controllability are first-order system objectives~\cite{reasoningforwireless}.

\section{Trajectory Shaping by Knowledge: Modulation Mechanisms and Cost Tradeoffs}
\label{sec:trajectory_shaping}

\begin{figure*}[t]
    \centering
    \includegraphics[width=0.98\textwidth]{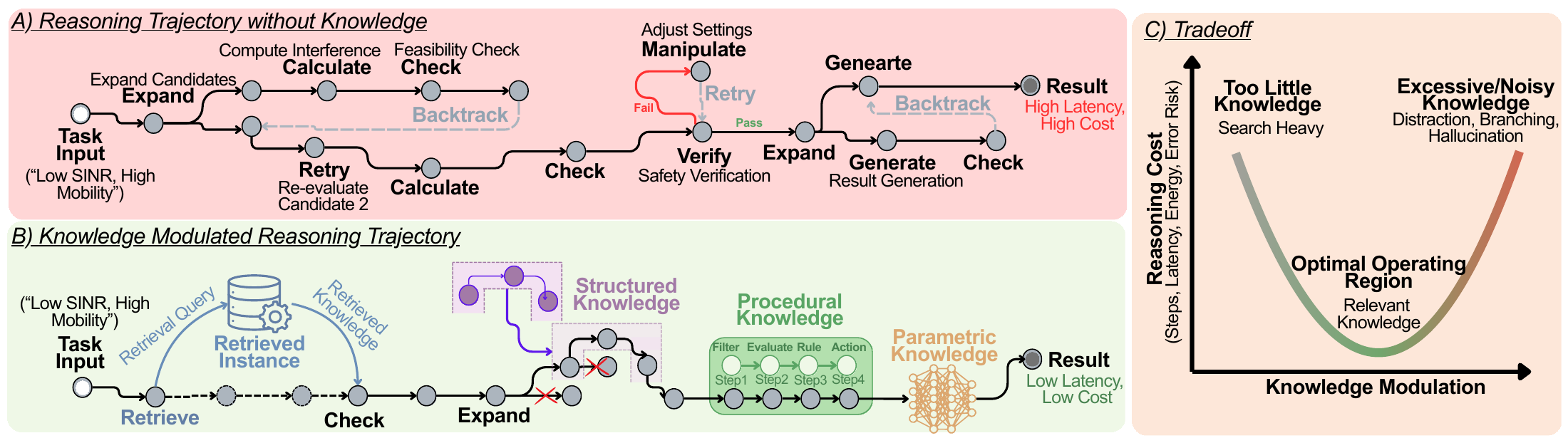}
    \caption{Knowledge modulation of reasoning trajectories and its non-monotonic tradeoff in a representative wireless mobility episode (e.g., handover under low SINR and high mobility). Panel A shows a search-dominated trajectory without effective knowledge, where the agent expands candidates, performs repeated calculation and feasibility checks, and backtracks with retries and safety verification, leading to high latency and high cost. Panel B shows knowledge-modulated trajectories, where retrieval knowledge enables a retrieve-and-adapt shortcut via a matched instance, structured knowledge prunes infeasible branches by enforcing constraints, procedural knowledge compresses reasoning into a deterministic workflow with predictable step budget, and parametric knowledge provides fast pattern completion when the regime is familiar, jointly yielding low latency and low cost. Panel C summarizes the non-monotonic tradeoff between knowledge modulation and reasoning cost (steps, latency, energy, and error risk): too little effective knowledge leads to search-heavy reasoning, whereas excessive or noisy knowledge induces distraction, branching, and hallucination, and an optimal operating region is achieved by relevant knowledge activation.}
    \label{fig:trajectory_tradeoff}
\end{figure*}

Having clarified how different forms of knowledge are represented and what reasoning capabilities they provide, we now examine how knowledge reshapes reasoning trajectories in practical agentic systems. An insight is that knowledge does not merely shorten a reasoning chain. Instead, it changes how the agent forms hypotheses, resolves uncertainty, and commits to actions. Consequently, knowledge affects trajectory length, branching behavior, and dominant failure modes, which are the main contributors to system cost under SWAP-C constraints. These effects are summarized in Fig.~\ref{fig:trajectory_tradeoff}, which contrasts a search-heavy trajectory without effective knowledge and several knowledge-modulated shortcuts.

\subsection{How Knowledge Modulates Reasoning Trajectories}
\label{subsec:knowledge_modulation}

Retrieval knowledge reshapes trajectories through instance reuse. When a highly aligned record is available, the agent can replace multi-step derivation with a short retrieve-and-adapt routine, and therefore collapse a large portion of intermediate reasoning. In wireless communication, this corresponds to retrieving past episodes with similar radio and mobility context and reusing the associated decisions as a strong prior for the current episode. This mechanism underlies retrieval-augmented reasoning systems. For example, the authors in~\cite{asai2024self} proposed Self-RAG, where the model decides when retrieval is necessary and critiques retrieved evidence before it is injected, which reduces reasoning depth and token consumption. Moreover,~\cite{mallen2023not} studied when a model should rely on internal signals versus external retrieval, and it showed that adaptive retrieval improves both efficiency and accuracy when the model's internal confidence is low, such as when the agent encounters an unfamiliar channel or mobility condition that was underrepresented during training. At the same time, retrieval introduces a relevance-sensitive failure mode. When instances are weakly aligned, the agent can be anchored to an inappropriate template and produce high-confidence errors. The authors in~\cite{dhuliawala2024chain} reported that misaligned evidence can induce such failures and proposed chain-based verification to filter unreliable inputs before they influence downstream steps. These results collectively indicate that retrieval enables the most aggressive trajectory collapse, while its reliability depends on relevance-aware selection and lightweight validation.

Structured knowledge reshapes trajectories by reorganizing reasoning around an explicit equation-level or program-level structure. By externalizing operator roles and admissible transformations, it fixes the decomposition order and stabilizes intermediate commitments, so that uncertainty is reduced before rule or tool selection. In wireless systems, such structure can be instantiated as explicit constraint graphs or factor-graph style formulations that encode interference coupling, feasibility constraints, and service requirements for scheduling and resource allocation. 
This mechanism appears in structure-aware reasoning systems that inject symbolic constraints into generation. For example,~\cite{banerjee2025crane} explored constrained decoding under symbolic restrictions to stabilize convergence, while~\cite{nagy2026chopchop} employed programmable abstract-syntax-tree constraints to reduce invalid traces. Importantly, structured knowledge does not eliminate exploration entirely. Instead, it changes where exploration happens by making intermediate commitments more stable and reducing early divergence.

Procedural knowledge reshapes trajectories by providing an executable workflow. After a suitable procedure is selected, the agent can transition from open-ended exploration to a short sequence of deterministic steps, which yields a predictable cost profile and improved controllability. In wireless mobility control, procedural knowledge often corresponds to standardized decision workflows, such as measurement filtering followed by candidate evaluation and rule checks (such as hysteresis and time-to-trigger) before committing to beam switching or handover actions~\cite{paulius2019survey}. This paradigm underlies program-aided and tool-guided reasoning, where reusable procedures replace long natural-language deliberation. For example,~\cite{gao2025efficient} demonstrated that generating executable programs can reduce reasoning depth and support execution-based verification while stabilizing inference latency when budgets are tight. 

Parametric knowledge reshapes trajectories through implicit pattern completion. When the input falls within the model's learned support, the agent can generate an output with minimal explicit intermediate structure, and thus achieve low latency without external memory access. In wireless communications, deep unfolding provides a canonical pathway to parametric knowledge by mapping iterative algorithmic solvers into trainable networks, such as unfolding projected-gradient style detectors for MIMO detection (such as DetNet) so that decisions can be produced with a fixed-depth forward pass~\cite{deka2025comprehensive}. At the same time, this pathway offers limited controllability because intermediate anchors for verification may not be present. Recent studies have examined how parametric knowledge interacts with external memory. For example,~\cite{xie2023adaptive} considered semi-parametric designs that combine internal representations with retrieval to balance latency and reliability. 

These four mechanisms are visualized together in Fig.~\ref{fig:trajectory_tradeoff} (Panel~B) using a representative wireless mobility episode (e.g., handover under low SINR and high mobility), where retrieval knowledge enables a retrieve-and-adapt shortcut, structured knowledge prunes infeasible branches, procedural knowledge compresses reasoning into a deterministic workflow with a predictable step budget, and parametric knowledge provides fast pattern completion when the regime is familiar.

In practice, these mechanisms are not mutually exclusive and often co-exist within a single reasoning episode. A common system-controlled pattern starts with a fast parametric hypothesis. Retrieval knowledge is then invoked if highly aligned instances can be found. If retrieval knowledge is unavailable or insufficient, structured knowledge is used to externalize task composition and admissible operations, which stabilizes subsequent decisions. Finally, procedural knowledge is activated once a suitable workflow is identified, so that the remaining reasoning can be executed through a short and deterministic sequence. Not every stage is triggered for every task. Simple inputs tend to be resolved by parametric shortcuts, whereas novel or complex cases progressively activate retrieval, structure, and procedures. This coordinated stack perspective is important for mobile settings, because activation order directly affects efficiency, reliability, and controllability under SWAP-C constraints. However, this layered activation also implies that each additional knowledge source carries both benefit and overhead. When activation is well-targeted, the stack converges quickly; when it is excessive or misaligned, it can introduce reconciliation cost and competing anchors that degrade rather than improve performance. This tension gives rise to the non-monotonic tradeoff that we analyze next.

\subsection{Reasoning as a Cost-Bearing Trajectory and the Non-Monotonic Tradeoff of Knowledge Activation}
\label{subsec:cost_tradeoff_activation}

In mobile agentic pipelines, reasoning is a cost-bearing trajectory that transforms task inputs into valid outputs through a sequence of intermediate steps~\cite{gao2025efficient}. In wireless-enabled agents, these intermediate states may further trigger measurement access, signaling exchange, or edge-cloud synchronization, and therefore directly translate into latency and energy overhead under tight budgets.

From a system perspective, the cost of reasoning mainly comes from three sources.
\begin{enumerate}
\item Token generation and compute cost, which scales with the number of intermediate steps and directly impacts latency and energy consumption.
\item Memory and history retention cost, which scales with how much intermediate state must be preserved for subsequent steps. Mobile agents rely on a limited input window and constrained on-device memory. As trajectories grow, the agent either truncates history, which harms reliability, or repeatedly retrieves and re-injects history, which increases communication overhead.
\item Branching and backtracking cost, which emerges under uncertainty. Without strong priors, the agent may explore multiple candidate paths, backtrack after failures, and perform repeated checks, so tokens and tool calls are spent on navigation rather than final outputs.
\end{enumerate}

Consequently, reasoning optimization on mobile platforms is naturally cast as trajectory shaping. Knowledge is a primary instrument for this shaping because it can inject priors, structure, and shortcuts into the reasoning loop. However, its effect is non-monotonic, since activation changes not only trajectory length but also branching behavior, as illustrated in Fig.~\ref{fig:trajectory_tradeoff} (Panel~C).

When effective knowledge is insufficient, the agent operates in a search-dominated regime. Reasoning relies on exploration, expansion, and backtracking, leading to long trajectories and high branching factors~\cite{gao2025efficient}. As a result, latency and energy consumption increase, and cumulative reasoning errors become more likely.

As effective knowledge activation increases, reasoning cost decreases rapidly. Relevant retrieval shortcuts can collapse portions of the trajectory~\cite{asai2024self}, structured constraints can prune infeasible branches~\cite{banerjee2025crane}, and procedural knowledge can convert exploration into a compact and more deterministic execution path~\cite{gao2025efficient}. In this region, knowledge shortens trajectories and reduces branching simultaneously, yielding improved efficiency and stability.

However, further increasing the quantity of knowledge can degrade performance. Weakly aligned retrieval may induce anchoring and high-confidence hallucinations~\cite{dhuliawala2024chain, mallen2023not}, while excessive or conflicting constraints may increase reconciliation overhead and trigger additional branching~\cite{nagy2026chopchop}. The agent may oscillate among competing strategies or commit to inappropriate templates, which increases both reasoning cost and error risk.

These observations indicate that the dominant system challenge is not knowledge storage, but knowledge activation control. The system must regulate which knowledge is exposed, when it is activated, and how its applicability is verified under tight communication and energy budgets. Accordingly, the objective of knowledge engineering for mobile agentic AI reasoning is not to maximize knowledge volume, but to optimize relevance-aware activation.

\section{Case Study: Knowledge-Driven AI Reasoning for Low-Altitude Wireless Networks}
\label{sec:case_study}

We examine knowledge-driven reasoning in a low-altitude UAV aerial base station mission. The UAV must repeatedly decide which ground user cluster to serve next, where to hover for service, and whether to request cross-episode knowledge under intermittent backhaul. In this scenario, we have to ensure that a service can meet a target rate, while the backhaul link gates when the UAV can synchronize knowledge from a Home ground station. In addition, runtime disruptions invalidate precomputed decisions, which makes reliable operation depend on on-device macro reasoning rather than one-shot cloud planning.

\subsection{Macro Reasoning With Deterministic Execution}
\begin{figure}[t]
\centering
\includegraphics[width=0.48\textwidth]{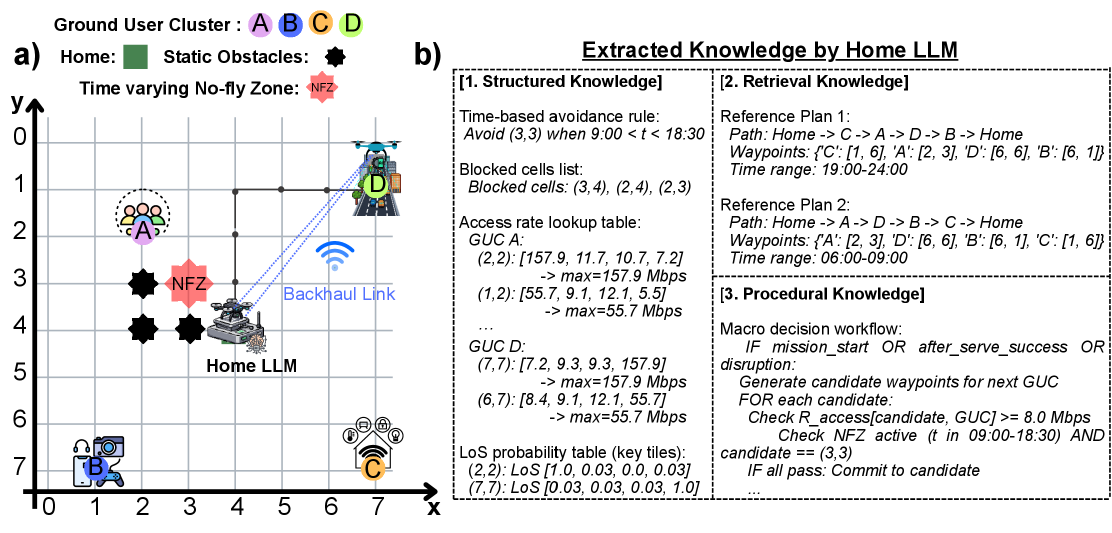}
\caption{Low-altitude UAV aerial base station case study and Home-promoted knowledge. (a) An 8 by 8 service area with Home, four ground user clusters, static obstacles, and a time-varying NFZ tile. The access link determines service feasibility, while intermittent backhaul determines when the UAV can synchronize knowledge from Home. (b) A compact knowledge pack promoted from historical mission logs at Home, including structured knowledge, retrieval knowledge, and procedural knowledge.}
\label{fig:cs_scenario_knowledge}
\end{figure}

Fig.~\ref{fig:cs_scenario_knowledge}(a) shows an 8 by 8 area where each cell is 500~m by 500~m. Home is located at cell (4,4), and four ground user clusters are placed near the corners. The UAV flies at a fixed altitude of 100~m. The access link uses 10~MHz bandwidth and 30~dBm transmit power. A cluster is considered served only if the access rate at the chosen hover waypoint exceeds 8~Mbps. The backhaul link uses 20~MHz bandwidth and 33~dBm transmit power. Knowledge transfer is feasible only when the backhaul rate exceeds 5~Mbps, which is the minimum rate required to deliver the knowledge pack within a single decision interval. This transfer is intermittent because the backhaul link quality varies with UAV-Home geometry and line-of-sight probability.

The environment contains three static obstacle tiles and one time-varying no-fly zone (NFZ) tile at (3,3). The NFZ location is known onboard, while its activation schedule is not provided as onboard data. The schedule is discoverable only from historical mission logs, and it is therefore promoted and stored at Home as knowledge.

We also model runtime disruptions that are only revealed during execution: a user cluster may drift by one tile upon arrival, or the measured SNR at a planned waypoint may drop below prediction due to temporary scattering or blockage, invalidating the current serve order and hover waypoint and triggering replanning.

Reasoning in this case study is macro-level replanning rather than low-level motion. At each decision point, the onboard model outputs a macro decision that includes the next cluster to serve, the hover waypoint, and whether to request knowledge when the backhaul permits. A deterministic motion module then executes the decision by moving the UAV to the chosen waypoint while avoiding obstacle tiles. This design keeps the reasoning trace interpretable and separates reasoning from path execution.

We measure reasoning cost by counting Reasoning Steps and Reasoning Tokens from the LLM trace, where a reasoning step is a marked macro reasoning step and is not a UAV movement step. Violation denotes constraint failure and includes NFZ entry during its active window, obstacle collision, and failed service attempts where the access rate at the selected hover waypoint does not meet the 8~Mbps threshold.

\subsection{Knowledge Promotion at Home and Experimental Setting}
Home promotes reusable structure offline from accumulated mission logs and stores it as a compact knowledge pack. In our implementation, we use Gemini 2.0 Flash at Home as the knowledge extractor and summarizer\footnote{\url{https://blog.google/innovation-and-ai/models-and-research/google-deepmind/google-gemini-ai-update-december-2024/}.}. 
This heterogeneous model assignment reflects a realistic deployment: the Home station has sufficient compute to run a capable extraction model, while the onboard agent is constrained to a lightweight model that fits within the UAV's SWAP-C budget. We instantiate the DIKW representation in this mission as follows. Data are per-decision signals available onboard, including UAV position, remaining clusters, remaining step budget, the local clock, and the last-action outcome label, such as moved, blocked, or served. Information is the episode trace, including visited tiles, chosen hover waypoints, measured access and backhaul outcomes, disruption indicators, and termination reasons. Knowledge is the persistent structure promoted from accumulated information and reused across episodes.

As shown in Fig.~\ref{fig:cs_scenario_knowledge}(b), the knowledge package contains three explicit formats. Structured knowledge externalizes feasibility and constraints, including the NFZ activation rule, the obstacle list, a backhaul feasibility map, and an access feasibility lookup for waypoint and user-cluster pairs. Retrieval knowledge stores a small set of successful reference mission plans, represented by serve orders and waypoint sequences with achieved throughput. Procedural knowledge provides a short serve-and-verify workflow, including access link check, backhaul check for knowledge requests, legality check, and commit or reject. These formats are designed to reshape the reasoning trajectory so that the onboard agent relies less on trial-and-error replanning and more on lookup, reuse, and deterministic verification.

For onboard macro reasoning, we use Qwen2.5-3B\footnote{\url{https://huggingface.co/Qwen/Qwen2.5-3B}.}. Qwen\_with\_k requests and caches the Home knowledge pack when the backhaul supports transfer, while Qwen\_no\_k runs the same macro reasoning loop without receiving the pack. To isolate parametric knowledge, we also run Gemini 2.0 Flash onboard without injecting the explicit knowledge pack, denoted as Gemini\_no\_k. This baseline captures stronger implicit priors that can improve decision quality, while remaining less externalizable and less budgetable than explicit promoted knowledge under SWAP-C constraints.

To provide a competitive cloud baseline, we include a cloud replanning method, denoted as home\_replan, implemented at Home using Gemini 2.0 Flash. Upon a disruption, the UAV uplinks a compact observation summary, Home replans, and then downlinks an updated macro plan. If the backhaul is in outage at the disruption moment, the closed loop is broken and the UAV must continue with a stale plan until a feasible checkpoint. 
We also include a non-LLM baseline, DRL\_PPO, which is a step-level policy that outputs micro movement actions and provides no reasoning trace. Unlike the LLM-based methods that operate at the macro decision level, PPO acts at the per-tile movement level, so its action granularity is finer and its reasoning cost is not directly comparable; we include it to benchmark mission completion and constraint satisfaction rather than reasoning efficiency. PPO is trained for 200,000 environment steps before evaluation.

Each method is evaluated over 50 inference episodes, and we report averages. Episode start time is randomized so that the NFZ boundary is not fixed, and disruptions follow the same sampling process across methods.

\begin{figure}[t]
\centering
\includegraphics[width=0.48\textwidth]{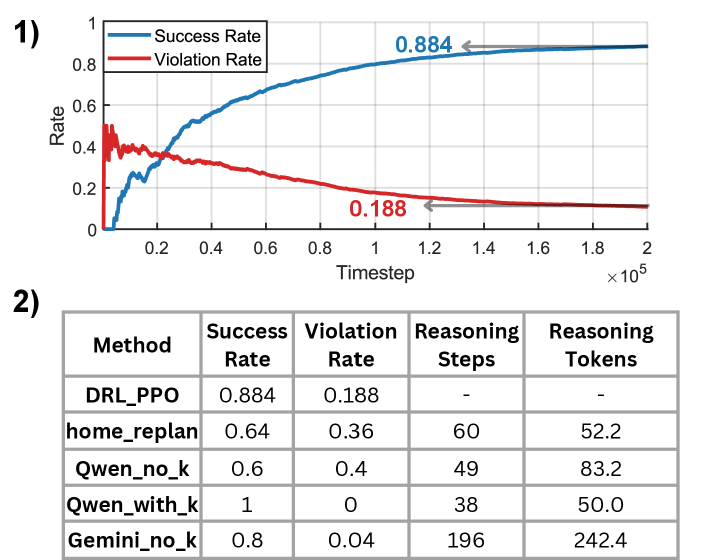}
\caption{Main results under intermittent backhaul and runtime disruptions. Panel 1) shows PPO training curves of success rate and violation rate over 200,000 environment steps. Panel 2) summarizes evaluation results averaged over 50 inference episodes per method. Violation aggregates NFZ violation, obstacle collision, and failed service attempts that do not meet the 8~Mbps access-rate target. Reasoning Steps and Reasoning Tokens are measured from the LLM macro reasoning trace, where a reasoning step is not a UAV movement step.}
\label{fig:cs_main_results}
\end{figure}

\subsection{Results and Analysis: Reliability, Reasoning Cost, and Wireless Tradeoffs}
Fig.~\ref{fig:cs_main_results} shows that DRL\_PPO reaches a success rate of 0.884 after training, yet its violation rate remains 0.188. This indicates that step-level control can learn to complete missions, but it is still brittle under latent time-varying constraints and wireless feasibility.

Without explicit knowledge, Qwen\_no\_k achieves lower reliability and higher violation. Its macro reasoning trace is also longer and more verbose, which is consistent with trial-and-error replanning when feasibility structure must be inferred on the fly. After enabling Home-promoted knowledge, Qwen\_with\_k achieves perfect reliability with zero violation, while reducing both the number of reasoning steps and the tokens consumed per reasoning step. This pattern matches the intended mechanism, since structured lookup and procedural verification prune infeasible options early, and retrieval references reduce search-heavy branching.

The Gemini\_no\_k baseline clarifies the role of parametric knowledge. It improves reliability relative to Qwen\_no\_k, which indicates that stronger implicit priors can partially compensate for missing explicit structure. However, it incurs substantially higher reasoning cost, with far more reasoning steps and much higher tokens per reasoning step. This gap suggests that parametric knowledge can help performance, yet it is less controllable and less cost efficient than explicit promoted knowledge when the agent must repeatedly re-decide under uncertainty.

The cloud replanning baseline highlights a wireless limitation that is orthogonal to compute. Although Home can replan with a stronger model, replanning depends on a timely uplink and downlink at the disruption moment. Backhaul outages break the closed loop exactly when a repair is needed, which explains why cloud replanning can be less reliable than on-device reasoning with cached knowledge under intermittent connectivity.

\begin{figure}[t]
\centering
\includegraphics[width=0.48\textwidth]{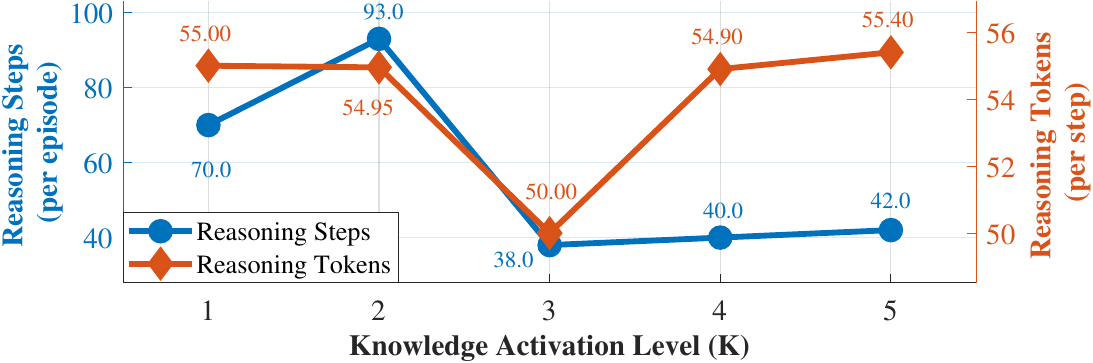}
\caption{Non-monotonic reasoning tradeoff when sweeping knowledge exposure level K for Qwen\_with\_k. Moderate exposure minimizes reasoning steps and tokens per reasoning step, while too little or too much knowledge increases reasoning cost due to search-heavy replanning and redundancy-induced reconciliation.}
\label{fig:cs_k_tradeoff}
\end{figure}

Finally, we vary the knowledge exposure level $K$ for Qwen\_with\_k, as shown in Fig.~\ref{fig:cs_k_tradeoff}. Each level incrementally adds content to the knowledge pack injected into the onboard prompt: $K{=}1$ includes only the core structured constraints (NFZ rule, obstacle list), $K{=}2$ adds the backhaul feasibility map, $K{=}3$ further adds the procedural serve-and-verify workflow and access feasibility lookup, $K{=}4$ adds retrieval reference plans, and $K{=}5$ includes the full pack with additional entries and extended annotations. The results confirm a non-monotonic tradeoff in reasoning cost. At moderate exposure ($K{=}3$), the agent achieves the fewest reasoning steps and the lowest tokens per step, since the knowledge pack provides sufficient feasibility structure and procedural guidance to support efficient decision making. At small $K$, the agent lacks critical constraints and reference plans, which forces it back into search-heavy replanning with longer and more verbose traces. At large $K$, repeated or weakly aligned entries introduce redundancy and reconciliation overhead, causing the reasoning trace to become more branchy and more token-intensive as the agent attempts to reconcile competing cues. Beyond reasoning cost, the sweep also reveals a wireless communication tradeoff. Increasing $K$ raises the one-time backhaul payload for knowledge synchronization from 125 tokens at $K{=}1$ to 1,234 tokens at $K{=}5$. Nevertheless, even the largest knowledge pack remains smaller than the 1,870-token closed-loop exchange required by home\_replan per disruption event. Moreover, the knowledge pack is cacheable and its synchronization cost can be amortized across all subsequent replanning episodes when the backhaul becomes unavailable, whereas cloud replanning must repeatedly pay the uplink-downlink cost whenever a new disruption occurs.

\section{Conclusion}
This paper presents knowledge-driven reasoning for mobile agentic AI. Through a DIKW-inspired taxonomy, we have considered knowledge as retrieval, structured, procedural, and parametric representations and shown that each reshapes reasoning trajectories differently in terms of branching, verification, and cost budgetability. An important finding is that knowledge exposure is non-monotonic: moderate, well-aligned activation minimizes reasoning cost, while both insufficient and excessive exposure degrade efficiency and reliability. The UAV aerial base station case study has confirmed that explicit, cacheable knowledge synchronized over intermittent backhaul outperforms knowledge-free on-device reasoning and cloud-centric replanning in both reliability and efficiency. These results motivate relevance-aware activation control and communication-aware knowledge packaging as first-class design principles for wireless-native agentic systems. Future directions include adaptive activation gating that jointly optimizes knowledge selection and wireless resource allocation, and online knowledge promotion from incremental mission experience.
\bibliographystyle{IEEEtran}
\bibliography{main}

@article{Survey_2512,
  title={Memory in the Age of AI Agents},
  author={Hu, Yuyang and Liu, Shichun and Yue, Yanwei and Zhang, Guibin and Liu, Boyang and Zhu, Fangyi and Lin, Jiahang and Guo, Honglin and Dou, Shihan and Xi, Zhiheng and others},
  journal={arXiv preprint arXiv:2512.13564},
  year={2025}
}

@article{memory_survey_agents_2025,
author = {Zhang, Zeyu and Dai, Quanyu and Bo, Xiaohe and Ma, Chen and Li, Rui and Chen, Xu and Zhu, Jieming and Dong, Zhenhua and Wen, Ji-Rong},
title = {A Survey on the Memory Mechanism of Large Language Model-based Agents},
year = {2025},
issue_date = {November 2025},
publisher = {Association for Computing Machinery},
address = {New York, NY, USA},
volume = {43},
number = {6},
issn = {1046-8188},
url = {https://doi.org/10.1145/3748302},
doi = {10.1145/3748302},
journal = {ACM Trans. Inf. Syst.},
month = sep,
articleno = {155},
numpages = {47}
}

@inproceedings{farahani_interplay_2024,
    title = "Deciphering the Interplay of Parametric and Non-parametric Memory in Retrieval-augmented Language Models",
    author = "Farahani, Mehrdad  and
      Johansson, Richard",
    editor = "Al-Onaizan, Yaser  and
      Bansal, Mohit  and
      Chen, Yun-Nung",
    booktitle = "Proceedings of the 2024 Conference on Empirical Methods in Natural Language Processing",
    month = nov,
    year = "2024",
    address = "Miami, Florida, USA",
    publisher = "Association for Computational Linguistics",
    url = "https://aclanthology.org/2024.emnlp-main.943/",
    doi = "10.18653/v1/2024.emnlp-main.943",
    pages = "16966--16977"
}

@article{rowley2007wisdom,
  title={The wisdom hierarchy: representations of the DIKW hierarchy},
  author={Rowley, Jennifer},
  journal={Journal of information science},
  volume={33},
  number={2},
  pages={163--180},
  year={2007},
  publisher={Sage Publications Sage CA: Thousand Oaks, CA}
}

@article{reasoningforwireless,
  title={Ai reasoning for wireless communications and networking: A survey and perspectives},
  author={Luo, Haoxiang and Yan, Yu and Bian, Yanhui and Feng, Wenjiao and Zhang, Ruichen and Liu, Yinqiu and Wang, Jiacheng and Sun, Gang and Niyato, Dusit and Yu, Hongfang and others},
  journal={arXiv preprint arXiv:2509.09193},
  year={2025}
}

@inproceedings{gao2025efficient,
  title={Efficient tool use with chain-of-abstraction reasoning},
  author={Gao, Silin and Dwivedi-Yu, Jane and Yu, Ping and Tan, Xiaoqing Ellen and Pasunuru, Ramakanth and Golovneva, Olga and Sinha, Koustuv and Celikyilmaz, Asli and Bosselut, Antoine and Wang, Tianlu},
  booktitle={Proceedings of the 31st International Conference on Computational Linguistics},
  pages={2727--2743},
  year={2025}
}

@inproceedings{asai2024self,
  title={Self-RAG: Learning to Retrieve, Generate, and Critique through Self-Reflection},
  author={Asai, Akari and Wu, Zeqiu and Wang, Yizhong and Sil, Avi and Hajishirzi, Hannaneh},
  booktitle={International Conference on Learning Representations},
  year={2024}
}

@article{banerjee2025crane,
  title={CRANE: Reasoning with constrained LLM generation},
  author={Banerjee, Debangshu and Suresh, Tarun and Ugare, Shubham and Misailovic, Sasa and Singh, Gagandeep},
  journal={arXiv preprint arXiv:2502.09061},
  year={2025}
}

@article{nagy2026chopchop,
  title={ChopChop: A Programmable Framework for Semantically Constraining the Output of Language Models},
  author={Nagy, Shaan and Zhou, Timothy and Polikarpova, Nadia and D'Antoni, Loris},
  journal={Proceedings of the ACM on Programming Languages},
  volume={10},
  number={POPL},
  pages={1905--1932},
  year={2026},
  publisher={ACM New York, NY, USA}
}

@article{han2025reasoning,
  title={Reasoning with Graphs: Structuring Implicit Knowledge to Enhance LLMs Reasoning},
  author={Han, Haoyu and Xie, Yaochen and Liu, Hui and Tang, Xianfeng and Nag, Sreyashi and Headden, William and Li, Yang and Luo, Chen and Ji, Shuiwang and He, Qi and others},
  journal={CoRR},
  year={2025}
}

@inproceedings{mallen2023not,
  title={When not to trust language models: Investigating effectiveness of parametric and non-parametric memories},
  author={Mallen, Alex and Asai, Akari and Zhong, Victor and Das, Rajarshi and Khashabi, Daniel and Hajishirzi, Hannaneh},
  booktitle={Proceedings of the 61st Annual Meeting of the Association for Computational Linguistics (Volume 1: Long Papers)},
  pages={9802--9822},
  year={2023}
}

@inproceedings{dhuliawala2024chain,
  title={Chain-of-verification reduces hallucination in large language models},
  author={Dhuliawala, Shehzaad and Komeili, Mojtaba and Xu, Jing and Raileanu, Roberta and Li, Xian and Celikyilmaz, Asli and Weston, Jason},
  booktitle={Findings of the association for computational linguistics: ACL 2024},
  pages={3563--3578},
  year={2024}
}

@inproceedings{xie2023adaptive,
  title={Adaptive chameleon or stubborn sloth: Revealing the behavior of large language models in knowledge conflicts},
  author={Xie, Jian and Zhang, Kai and Chen, Jiangjie and Lou, Renze and Su, Yu},
  booktitle={The Twelfth International Conference on Learning Representations},
  year={2023}
}

@article{paulius2019survey,
  title={A survey of knowledge representation in service robotics},
  author={Paulius, David and Sun, Yu},
  journal={Robotics and Autonomous Systems},
  volume={118},
  pages={13--30},
  year={2019},
  publisher={Elsevier}
}

@article{deka2025comprehensive,
  title={Comprehensive Review of Deep Unfolding Techniques for Next-Generation Wireless Communication Systems},
  author={Deka, Sukanya and Deka, Kuntal and Nguyen, Nhan Thanh and Sharma, Sanjeev and Bhatia, Vimal and Rajatheva, Nandana},
  journal={arXiv preprint arXiv:2502.05952},
  year={2025}
}
\end{document}